\documentclass[12pt]{article}

\usepackage{amsmath, amssymb}
\usepackage{mathrsfs}
\usepackage{physics}
\usepackage[T1]{fontenc}                                                                                                                                                                                                                                                                                                                                                                                                                                                                                                                       
\usepackage{fancyhdr}
\usepackage{appendix}
\usepackage[pdftex]{graphicx}
\usepackage{titlesec}
\usepackage{cite}
\usepackage[top=25truemm,bottom=25truemm,left=25truemm,right=25truemm]{geometry}

\DeclareFontShape{T1}{cmr}{mx}{n}{<->cmr10}{}

\titleformat*{\section}{\large\bfseries}
\titleformat*{\subsection}{\normalsize\bfseries}

\newcommand{\beq}{\begin{equation}}
\newcommand{\eeq}{\end{equation}}
\newcommand{\beqn}{\begin{equation*}}
\newcommand{\eeqn}{\end{equation*}}

\newcommand\als[1]{\begin{align}\begin{split}#1\end{split}\end{align}}
\newcommand{\im}{\mathrm{i}}

\makeatletter

\@addtoreset{equation}{section}
\makeatother

\begin{document}

\fontseries{mx}\selectfont

\begin{titlepage}
\renewcommand{\thefootnote}{\fnsymbol{footnote}}
\begin{normalsize}
\begin{flushright}
\begin{tabular}{l}
UTHEP-734
\end{tabular}
\end{flushright}
  \end{normalsize}

~~\\

\vspace*{0cm}
    \begin{Large}
       \begin{center}
         {Diffeomorphisms on Fuzzy Sphere}
       \end{center}
    \end{Large}
\vspace{0.7cm}

\begin{center}
Goro I\textsc{shiki}$^{1),2)}$\footnote[0]
            {
e-mail address : 
ishiki@het.ph.tsukuba.ac.jp} and
Takaki M\textsc{atsumoto}$^{2),3)}$\footnote[0]
            {
e-mail address : 
matsumoto@het.ph.tsukuba.ac.jp}

\vspace{0.7cm}

     $^{ 1)}$ {\it Tomonaga Center for the History of the Universe, University of Tsukuba, }\\
               {\it Tsukuba, Ibaraki 305-8571, Japan}\\
                   
     $^{ 2)}$ {\it Graduate School of Pure and Applied Sciences, University of Tsukuba, }\\
               {\it Tsukuba, Ibaraki 305-8571, Japan}\\
               
     $^{ 3)}$ {\it School of Theoretical Physics, Dublin Institute for Advanced Studies, }\\
               {\it 10 Burlington Road, Dublin 4, Ireland}\\
               \end{center}

\vspace{0.5cm}

\begin{abstract}
Diffeomorphisms can be seen as 
automorphisms of the algebra of functions. 
In the matrix regularization, functions on a smooth compact 
manifold are mapped to finite size matrices. 
We consider how diffeomorphisms  
act on the configuration space of the matrices through the 
matrix regularization. 
For the case of the fuzzy $S^2$, we construct the matrix regularization 
in terms of the Berezin-Toeplitz quantization.
By using this quantization map, we define diffeomorphisms on the space of 
matrices. 
We explicitly construct the matrix version of holomorphic 
diffeomorphisms on $S^2$.
We also propose three methods of constructing 
approximate invariants on the fuzzy $S^2$.
These invariants are exactly invariant under area-preserving diffeomorphisms 
and only approximately invariant (i.e. invariant in the large-$N$ limit) 
under the general diffeomorphisms.

\end{abstract}

\end{titlepage}

\tableofcontents

\setcounter{footnote}{0}
\section{Introduction}

The matrix regularization \cite{Hoppe;1982, deWit:1988wri}
gives a regularization of the world volume theory of membranes
with the world volume $\mathbf{R}\times \Sigma$
where $\Sigma$ is a compact Riemann surface with a fixed topology.
Although the original world volume theory has
the world volume diffeomorphism symmetry,
it is restricted to area-preserving diffeomorphisms on $\Sigma$
in the light-cone gauge.
In this gauge fixing, we have a Poisson bracket defined by a volume form on $\Sigma$,
which is invariant under the residual gauge transformations.
The matrix regularization is an operation of replacing
the Poisson algebra of  functions on $\Sigma$
by the Lie algebra of $N\times N$ matrices.
After this replacement, the world volume theory in the light-cone gauge becomes
a quantum mechanical system with matrix variables.
Remarkably enough, the regularized theory coincides with
the BFSS matrix model which is conjectured to give a complete 
formulation of M-theory
in the infinite momentum frame \cite{Banks:1996vh}.
This coincidence suggests that the matrix regularization is
not just a regularization of the world volume theory
but a fundamental formulation of M-theory.
The matrix regularization is also applied to
type IIB string theory and provides a matrix model for 
a nonperturbative formulation of the string theory \cite{Ishibashi:1996xs}.

The regularized membrane theory has the $U(N)$ gauge symmetry
which acts on the matrix variables as unitary similarity transformations.
This symmetry should correspond to
the are-preserving diffeomorphisms on $\Sigma$.
However, we have not completely understood
how general diffeomorphisms on $\Sigma$ act on the matrix 
variables\footnote{
In \cite{Hanada:2005vr}, it is shown that diffeomorphisms can be embedded into
the unitary transformations, if one considers the matrices as 
covariant derivative acting on an infinite dimensional Hilbert space.
This formulation is different from the matrix regularization, which 
we discuss in this paper. }.
Since diffeomorphisms should be essential in
constructing a covariant formulation of M-theory,
it is important to clarify the full diffeomorphisms in the matrix model.
The description of general diffeomorphisms in terms of matrices
may also enable us to formulate theories of gravity
on noncommutative spaces
\cite{Chamseddine:1996zu,Aschieri:2005zs,Steinacker:2010rh,Nair:2015rsa}
using the matrix regularization.

\begin{figure}
	\centering
	\includegraphics[width=10cm]{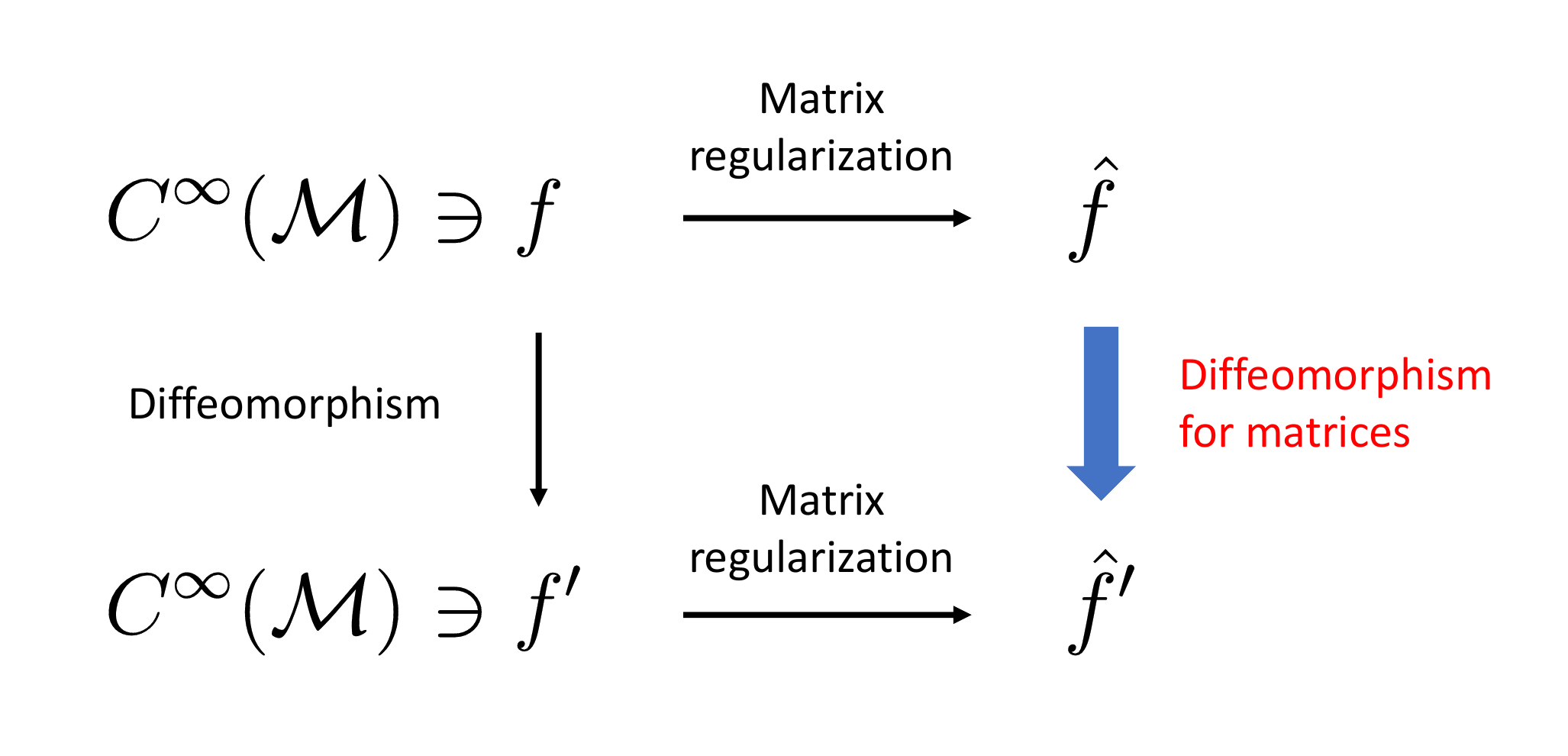}
	\caption
	{Under a diffeomorphism, a function $f$ is mapped to another 
function $f'$. These functions are then mapped to matrices by the 
matrix regularization. By comparing the two matrices, 
we can read off how the diffeomorphism acts on the space of the matrices.}
\label{fig1}
\end{figure}

In this paper, we focus on automorphisms of $C^\infty(\Sigma)$
induced by diffeomorphisms on $\Sigma$ rather than
diffeomorphisms themselves.
This is reasonable since 
the group of diffeomorphisms on $\Sigma$ is
isomorphic to automorphisms of $C^\infty(\Sigma)$.
Under the matrix regularization, automorphisms of $C^\infty(\Sigma)$ 
are mapped to transformations between matrices. 
See Fig.~\ref{fig1}. From this correspondence, we study 
how diffeomorphisms act on the space of the matrices.

For this purpose, we need to fix the scheme of the matrix regularization.
A systematic scheme is given by the Berezin-Toeplitz quantization
\cite{Klimek:1992,Bordemann:1993zv,Ma:2008}\footnote{
The same construction was also considered in the context of 
the tachyon condensation on D-branes
 \cite{Asakawa:2001vm, Terashima:2005ic, Terashima:2018tyi}
(See also \cite{Ellwood:2005yz}). 
This method is also related to 
the lowest Landau level problem \cite{Hasebe:2015htd,Hasebe:2017myo}. 
},
which is 
based on the concept of coherent states and has been
developed in the context of the geometric and the deformation quantizations.
In this paper, we construct the matrix regularization of $S^2$
in terms of the Berezin-Toeplitz quantization
and investigate how diffeomorphisms on $S^2$,
which are not necessarily are-preserving, act on the configuration space of 
the matrices.
In particular, for holomorphic diffeomorphisms on $S^2$,
we explicitly construct one-parameter deformations of 
the standard fuzzy $S^2$.
We also propose three kinds of 
approximate diffeomorphism invariants on the fuzzy $S^2$.
These are exactly invariant under area-preserving diffeomorphisms 
(the unitary similarity transformations)
and also invariant under general diffeomorphisms
in the large-$N$ limit.

The organization of this paper is as follows.
In section 2, we introduce the basic terminology and notation
concerning diffeomorphisms of a smooth manifold equipped with
geometric structures.
In section 3, we review the Berezin-Toeplitz quantization.
In section 4, we define the action of diffeomorphisms on the space of matrices.
In section 5, we construct the matrix regularization of $S^2$
based on the Berezin-Toeplitz quantization.
Then, we investigate the holomorphic diffeomorphisms for matrices.
In section 6, we propose the approximate invariants. 
In section 7, we summarize our results.

\section{Diffeomorphisms and automorphisms}
\label{section diff}

In this section, we review the notion of
diffeomorphisms preserving geometric structures.
See e.g. \cite{book} for more details.

Let $M$ be a smooth compact manifold.
We denote by $\text{Diff}(M)$ the group of diffeomorphisms from $M$ to itself\footnote{Recall that a differentiable 
map $\varphi:M\rightarrow M$ is called a diffeomorphism
if $\varphi$ is a bijection and its inverse is also differentiable.}.
Let $\varphi\in \text{Diff}(M)$.
For a smooth function $f$ on $M$,
$\varphi$ induces a new function on $M$ defined by
\beq
	\label{pull of f}
	f':=\varphi^\ast f=f\circ\varphi,
\eeq
where $\varphi^\ast$ is the pullback by $\varphi$.
The map $f\mapsto f'$ defines an automorphism of $C^\infty(M)$.
Inversely, an arbitrary automorphism of $C^\infty(M)$ is expressed
in the form (\ref{pull of f}) using a diffeomorphism.
This means that $\text{Diff}(M)$ is isomorphic to
the group of automorphisms of $C^\infty(M)$
\footnote{See e.g. Section 1.3 in \cite{GraciaBondia:2001tr} for a precise proof.}.
More generally, for a tensor field $T$ on $M$,
$\varphi$ induces a new tensor field $T'$ on $M$
as the pullback or the pushforward.
The map $T\mapsto T'$ does not change the type of $T$
but generally changes the components of $T$.
If $T=T'$, then we say that $\varphi$ preserves $T$.

Let $\{\varphi_t\}_{t\in \mathbf{R}}$ be a one parameter group of diffeomorphisms,
that is, the map from $\mathbf{R}\times M$ to $M$
defined by $(t,p)\mapsto\varphi_t(p)$ is smooth,
$\varphi_t\circ\varphi_s=\varphi_{t+s}$ for any $t,s\in\mathbf{R}$
and $\varphi_0=\text{id}_M$.
Since $\{\varphi_t\}_{t\in \mathbf{R}}$ gives a smooth curve
$t\mapsto\varphi_t(p)$ on $M$,
we can define the velocity vector field $u$ by
\beq
	\label{diff vector}
	(uf)(p)=\eval{\frac{d}{dt}f(\varphi_t(p))}_{t=0}.
\eeq
The infinitesimal transformation of $T$ induced by $\varphi_t$ is
\beq
	\label{infinitesimal diff}
	\delta T:=\lim_{t\to0}\frac{1}{t}(T'-T)=L_uT,
\eeq
where $L_u$ is the Lie derivative along $u$.
If and only if $L_uT=0$,
$\varphi_t$ preserves $T$ for any $t$.

We suppose that $T$ is a geometric structure on $M$,
that is, $T$ has some special properties.
For example, a Riemannian structure $g$ is
a positive definite symmetric tensor field of type $(0,2)$,
a symplectic structure $\omega$ is
a non-degenerate, closed antisymmetric tensor field of type $(0,2)$,
and a complex structure $J$ is
a tensor field of type $(1,1)$ satisfying $J\circ J=-\text{id}_M$ and 
the integrability condition. If $\varphi$ preserves $T$, 
then $\varphi$ is called an automorphism of $(M,T)$.
The subgroup of $\text{Diff}(M)$ consisting of
all automorphisms of $(M,T)$ is called
the automorphism group of $(M,T)$ and denoted by $\text{Aut}(M,T)$.
Similarly, if $\varphi$ preserves several structures $T, T', \ldots$, 
the corresponding automorphism group is denoted by 
$\text{Aut}(M,T, T', \ldots)$.

The automorphism groups $\text{Aut}(M,g)$, $\text{Aut}(M,\omega)$ and 
$\text{Aut}(M,J)$ are also known as the groups of isometries, 
symplectomorphisms
and holomorphic diffeomorphisms, respectively.
Automorphism groups are often isomorphic to a finite dimensional Lie group
depending on $T$ although $\text{Diff}(M)$ is an infinite dimensional 
Lie group.
For example, 
any isometry group is known to be
isomorphic to a finite dimensional Lie group.

For symplectic manifolds $(M, \omega)$ 
with the trivial first cohomology class, 
any vector field (\ref{diff vector}) generated by 
a symplectomorphism is a Hamiltonian vector field $u_{\alpha}$, 
which satisfies 
$d\alpha = \omega (u_\alpha , \,\cdot\,)$ with a function $\alpha$
on $M$.
Inversely, for any function $\alpha$, there is a
unique Hamiltonian vector field $u_\alpha$. Hence, the generators of 
symplectomorphisms are labelled by functions on $M$.
The infinitesimal transformation of a function $f$ induced by 
a symplectomorphism can be written as
\begin{align}
	\delta f =\omega(u_f, u_\alpha)= \{f,\alpha\},
	\label{area preserving for f}
\end{align}
where $\{\,\cdot\, ,\,\cdot\,\}$ is the Poisson bracket.
Since Hamiltonian vector fields satisfy 
$[u_\alpha, u_\beta]=u_{\{\alpha, \beta\}}$, 
the Lie algebra of $\text{Aut}(M,\omega)$ is isomorphic to 
the Poisson algebra on $M$, which is an infinite dimensional 
Lie algebra.

\section{Matrix regularization and Berezin-Toeplitz quantization}
\label{section 3}

In this section, we review the construction of the matrix regularization
based on the Berezin-Toeplitz quantization. 
In the following, we denote by $\{\,\cdot\, ,\,\cdot\,\}$
the poisson bracket induced by the symplectic form $\omega$.
We assume $(M, \omega)$ to be 
a $2n$-dimensional closed symplectic manifold.

Let $N_1,N_2,\ldots$ be a strictly monotonically increasing 
sequence of positive integers. 
The matrix regularization is formally defined by a family of linear maps
from functions on $(M, \omega)$ to $N_p\times N_p$ matrices,
$\{T_p: C^{\infty}(M)\rightarrow M_{N_p}(\mathbf C) \}_{p\in\mathbf{N}}$, 
which satisfy 
\als
{
	\label{MR}
	&\lim_{p\to\infty}\|T_p(f)T_p(g)-T_p(fg)\|=0,\\
	&\lim_{p\to\infty}\|p[T_p(f),T_p(g)]-\im T_p(\{f,g\})\|=0,
}
for any $f,g\in C^\infty(\Sigma)$ \cite{Arnlind:2010ac}. 
Here, $\|\cdot\|$ denotes an arbitrary matrix norm.
In order to avoid the trivial case with $T_p(f)=0$, one may also assume 
for example that 
$\lim_{p\to\infty}\Tr T_p(f) = \int_{M} \omega^n f / n!$.

The conditions (\ref{MR}) and the linearity of $T_p$ means that
$T_p$ is approximately a representation of the
Poisson algebra on $\mathbf{C}^{N_p}$. 
Note that the matrix algebra is noncommutative
and hence is never homomorphic to the commutative algebra of functions. 
The matrix regularization gives only an approximate homomorphism
and the accuracy of the approximation improves
as the matrix size tends to infinity.

The matrix regularization is closely related to the quantization of classical mechanics. Recall that, in the quantization, classical observables 
${\cal O}(q,p)$, which are functions on the phase space, 
are promoted to quantum operators $\hat{\cal O}(\hat{q}, \hat{p})$ and
the classical Poisson bracket is replaced with the commutator of the operators.
This relation is very similar to (\ref{MR}), where the large-$p$ limit 
in (\ref{MR}) corresponds to the classical limit $\hbar \rightarrow 0$.
However, there is a crucial difference. 
The Hilbert space for quantum mechanics is infinite dimensional, 
while that of the matrix regularization is finite dimensional.
This difference comes from the noncompactness of the classical phase space
(i.e. one needs infinitely many wave packets to cover the 
entire noncompact phase space. This would not be the case if the 
phase space were compact.).
In the matrix regularization, we always assume that the manifold $M$ is 
compact, so that the associated Hilbert space is finite dimensional.
Hence, the matrix regularization is said to be 
the quantization on a compact phase space.

The quantization of classical mechanics is essentially given by 
fixing the ordering of the operators. For the anti-normal ordering, 
the quantization can be elegantly reformulated as 
the Berezin-Toeplitz quantization, which has been
developed in the context of the geometric and the deformation quantizations
\cite{Klimek:1992,Bordemann:1993zv,Ma:2008}.
See appendix A for the Berezin-Toeplitz quantization for 
quantum mechanics.  
The Berezin-Toeplitz quantization has a great advantage that 
it can be applied not only to the flat space but also 
to a large class of manifolds with $\text{spin}^c$ structures,
giving a systematic way of generating the matrix regularizations
for compact $\text{spin}^c$ manifolds.

Let us review the Berezin-Toeplitz quantization for 
$(M,\omega)$. Our setup is as follows.
We choose a Riemannian metric $g$ and an almost complex 
structure $J$ such that they are compatible with $\omega$.
Then, $M$ has a $\text{spin}^c$ structure associated with $J$. 
For the moment, we assume that this gives a $\text{spin}$ structure.
The case of general $\text{spin}^c$ manifolds will be mentioned in the 
last part of this section.
Let $S$ be a spinor bundle over $M$.
The fiber of $S$ is a spinor space $W\cong\mathbf{C}^{2^n}$,
and spinor fields on $M$ are sections of $S$.
Let $P$ be a principle $U(1)$-bundle over $M$ with a gauge connection $A$
and the curvature two-form $F=dA$. 
We consider the case with $F=2\pi \omega V_n^{-1/n}$,
where $V_n=\int_{M} \omega^n/n!$ is the symplectic volume, 
so that $A$ is proportional to the symplectic potential\footnote{
This choice of $F$ is always possible for $n=1$. 
For $n \ge 2$, this is possible when 
$2\pi \omega V_n^{-1/n}$ belongs to 
the integer cohomology class. Such a manifold is called
a quantizable manifold in mathematical literatures
\cite{Bordemann:1993zv,Ma:2008,book2}.}. 
Let $L_p$ be an associated complex line bundle to $P$ for the
irreducible representation $\pi_p:U(1)\to GL(1,\mathbf{C})$ defined by
$\pi_p(e^{\im\theta})=e^{\im p \theta}$ ($\theta\in\mathbf{R}$, $p \in\mathbf{N}$).
We consider a twisted spinor bundle
$S\otimes L_p \simeq S\otimes L^{\otimes p}_1$ over $M$,
where $L^{\otimes p}_1$ stands for the $p$-fold tensor product of $L_1$.
The sections of this bundle are spinor fields on $M$
which take values in the representation space of $\pi_p$.
We denote by $\Gamma(S\otimes L_p)$
the vector space of the spinor fields and define a inner product by
\beq
	\label{inner product of spinor}
	(\psi,\phi)=\frac{1}{n!}\int_M
	\omega^n\, \psi^\dagger \cdot \phi ,
\eeq
for $\psi,\phi\in\Gamma(S\otimes L_p)$,
where $\psi^\dagger \cdot \phi $ denotes the Hermitian inner product on $W$
(i.e. the contraction of the spinor indices).

Then, we define the Dirac operator on $\Gamma(S\otimes L_p)$.
Let $U\subset M$ be an open subset
and $\sigma^\mu$ $(\mu=1,2,\ldots,2n )$ local coordinates on $U$.
We denote by $e_a$  $(a=1,2,\ldots,2n)$ an orthonormal frame on $U$
with respect to $g$ and by $\theta^a$ the dual basis of $e_a$. 
In the following, we raise and lower the indices of the orthonormal frame 
by using the Kronecker delta.
Now we define a linear map $\gamma$ from vector fields on $U$
to endomorphisms of $W$ by
\beq
	\label{linear map of e}
	\gamma(e_a)=\gamma_a,
\eeq
where $\gamma_a$ are the gamma matrices
satisfying $\{\gamma_a,\gamma_b\}=2\delta_{ab}\textbf{1}_{2^n}$. 
Using $\gamma_a$, we define the spin connection 
$\Omega^{ab}\gamma_a\gamma_b/4$,
where $\Omega_{ab}$ is a local one-form
determined by
\als
{
	\label{spin connection}
	&\Omega_{ab}+\Omega_{ba}=0,\\
	&\Omega^a{}_b\wedge \theta^b+d\theta^a=0.
}
Given these data, we define the Dirac operator on $\Gamma(S\otimes L_p)$ by
\beq
	\label{Dirac operator}
	D=\im\gamma(\partial^\mu)\biggl(\partial_\mu
	+\frac{1}{4}\Omega^{ab}_\mu\gamma_a\gamma_b-\im pA_\mu\biggr),
\eeq
where $\partial_\mu = {\partial}/{\partial \sigma^\mu}$.

Finally, we construct the quantization map 
satisfying (\ref{MR}). Let $\psi_i$ $(i=1,2,\ldots,N_p)$, be the orthonormal basis of $\text{Ker}D$ with respect to the inner product (\ref{inner product of spinor}), where $N_p=\text{dim}\text{Ker}D$. At least for large $p$, 
the sequence $N_p,N_{p+1},\ldots$ is in fact 
strictly monotonically increasing, 
as shown in appendix~\ref{Estimation of dimkerD}. 
We define the so-called Toeplitz operator by
\beq
	\label{Toeplitz operator}
	\mel*{i}{T_p(f)}{j}=(\psi_j,f\psi_i),
\eeq
for $f\in C^\infty(M)$, where $\{\ket*{i} \mid i=1,2,\cdots, N_p\}$ is
an orthonormal basis of $\textbf{C}^{N_p}$ corresponding to $\psi_i$. This is a 
generalization of (\ref{BT map new}).
In this construction, the map $T_p(f)$ indeed satisfies
the conditions (\ref{MR}) because of the asymptotic expansion \cite{Ma:2008},  
\beq
	\label{asymptotic expansion}
	T_p(f)T_p(g)= T_p(C_0(f,g))+ \frac{1}{p}T_p(C_1(f,g))+O(p^{-2})
\eeq
for any $f,g\in C^\infty(M) $, where
$C_0(f,g)=fg$ and $C_1(f,g)-C_1(g,f)=\im\{f,g\}$.

So far, we have assumed that $M$ has a spin structure. However, 
the similar construction is also available 
for general $\text{spin}^c$ manifolds.
In this case, an additional $U(1)$ connection is 
needed in the definition of the Dirac operator 
(\ref{Dirac operator}).

\section{Matrix diffeomorphisms}

In this section, we define the action of diffeomorphisms in
the configuration space of matrices using the Toeplitz operator.

Let $(M,\omega)$ be a closed symplectic manifold.
For $f\in C^\infty(M)$, we consider an automorphism
$f\mapsto \varphi^\ast f$ induced by $\varphi\in \text{Diff}\,(M)$.
By following the procedure in Fig. 1,
we define a transformation of $N_p\times N_p$ matrices by
\beq
 	\label{diff of Mat}
	T_p(f)\mapsto T_p(\varphi^\ast f).
\eeq
We call this transformation
a matrix diffeomorphism corresponding to $\varphi$.

It is well-known that area-preserving diffeomorphisms 
(\ref{area preserving for f})
are realized 
as unitary similarity transformations in the matrix regularization.
This can also be seen by comparing the symmetries of the 
light-cone membrane and the matrix model.
The definition (\ref{diff of Mat}) also realizes this correspondence.
From (\ref{asymptotic expansion}), one can see that 
the transformation (\ref{area preserving for f})
is mapped to the infinitesimal matrix diffeomorphism,
\begin{align}
\delta T_p(f) =  T_p(\delta f)= -\im p[T_p(f), T_p(\alpha)]+ O(p^{-1}).
\label{matrix ap-diffeo}
\end{align}
This is nothing but the infinitesimal form of a 
unitary similarity transformation.

Conversely, if (\ref{matrix ap-diffeo}) holds, 
then $\delta f$ is an area-preserving diffeomorphism.
This is shown as follows.
 Suppose that (\ref{matrix ap-diffeo}) holds
for a certain $\alpha \in C^{\infty}(M)$.
Then, because of (\ref{MR}), we have 
$T_p(\delta f - \{\alpha, f \})=O(p^{-1})$.
This is satisfied if and only if $\delta f - \{\alpha, f \}=0$
\cite{Arnlind:2010ac}.
Hence, $\delta f$ is area-preserving.
These arguments show that 
for non-area-preserving diffeomorphisms, the corresponding 
matrix diffeomorphisms cannot be written in the form 
(\ref{matrix ap-diffeo}).

Recall that diffeomorphisms can be regarded as
 automorphisms on the space of functions.
On the other hand, 
matrix diffeomorphisms are not necessarily
an automorphism of $M_{N_p}(\textbf{C})$,
which can always be written as a similarity transformation. 
This is because the Toeplitz operator is not an isomorphism
from $C^\infty(M)$ to $M_{N_p}(\textbf{C})$.
In fact, the definition (\ref{diff of Mat}) contains a much broader 
class of transformations than the similarity transformations.
In the next section, we will explicitly construct some of those 
transformations for fuzzy $S^2$.

\section{Matrix diffeomorphisms on fuzzy sphere}

In this section, we consider the Berezin-Toeplitz quantization 
and matrix diffeomorphisms on the fuzzy $S^2$
\cite{Madore:1991bw}.
We will explicitly construct holomorphic matrix diffeomorphisms
on the fuzzy $S^2$ and see that most of these transformations can not 
be written as a similarity transformation.

\subsection{Berezin-Toeplitz quantization on $S^2$}

We first construct the Berezin-Toeplitz quantization map for $S^2$.
See appendix~\ref{appendix geometric structure} for our notation and 
geometric structures on $S^2$, which we use below.

In the Berezin-Toeplitz quantization, 
we need spinors, which are sections of $S\otimes L_p$.
Here, we take the 
Wu-Yang monopole configuration (\ref{Wu-Yang}) 
as a connection of the line bundle $L_1$
and
$S$ is the bundle of two-component spinors.
The Dirac operator (\ref{Dirac operator}) 
on $\Gamma(S\otimes L_p)$ can be decomposed 
as (\ref{decomposition of D}). The local form of 
$D^{\pm}$ on $U_z$ are given as
\als
{
	D^+&=\sqrt{2}\im\left\{(1+|z|^2)\partial_{\bar z}+\frac{p-1}{2}z\right\}, \\
	D^-&=\sqrt{2}\im\left\{(1+|z|^2)\partial_z-\frac{p+1}{2}\bar z\right\}.
}
Here, we have used the geometric structures shown in
appendix~\ref{appendix geometric structure}.

In order to construct Toeplitz operators,
we need the zero modes of $D^\pm$.
We can easily solve the eigenvalue equations $D^\pm \psi^\pm=0$
and obtain $\psi^+=(1+|z|^2)^{-(p-1)/2}h^+$ and 
$\psi^-=(1+|z|^2)^{(p+1)/2}h^-$,
where $h^+$ and $h^-$ are arbitrary holomorphic
and anti-holomorphic functions on $U_z$, respectively.
Note that  the integral,
\beq
	\int_{S^2}\omega\,|\psi^-|^2
	=\im \int_{S^2}dz d\bar z\,(1+|z|^2)^{(p-1)}|h^-|^2,
\eeq
does not converge for $p\ge 1$,
 unless $h^-=0$.
Thus, we find that $\text{Ker}D^-=\{0\}$
for $p\ge 1$.
The similar integral for $\psi^+$ converges
when the degree of $h^+$ is smaller than $p$.
Such $h^+$ is a holomorphic polynomial of degree $p-1$,
which can be expanded in terms of the basis $1,z,z^2,\ldots,z^{p-1}$.
Therefore, we find that\footnote{
Note that these results are consistent with 
the vanishing theorem and the index theorem, 
$\text{Ind}D=p$.}
 $N_p=\text{dim}\text{Ker}D^+=p$.
The Dirac zero modes can be written as 
\beq
	\psi_i(z,\bar z)=\sqrt{\frac{p}{2\pi}}
	\left(\begin{array}{c}\braket*{i}{z}\\0\end{array}\right),
\eeq
where $\{\ket*{i} \mid i=1,2,\cdots, p \}$ is an arbitrary 
orthonormal basis of $\textbf{C}^p$,
and $\ket{z}$ is the Bloch coherent state with $J=(p-1)/2$
defined by
\beq
	\ket*{z}
	=\frac{1}{(1+|z|^2)^J}\sum^J_{r=-J}z^{J-r}
	{\binom{2J}{J+r}}^{1/2}\ket*{Jr}.
\label{def of bloch}
\eeq
Here, $\{\ket*{Jr} \mid r=-J,-J+1,\ldots,J \}$ is the standard basis
of the $(2J+1)$-dimensional irreducible representation space 
of $SU(2)$.
By using the resolution of identity, $p\int_{S^2}\omega\op*{z}/2\pi=\textbf{1}_p$, 
one can check that $\{\psi_i \mid i=1,2, \cdots, p \}$ is an 
orthonormal basis of $\text{Ker}D$.

In the above setup, the Toeplitz operators
 (\ref{Toeplitz operator}) are written as
\beq
	\label{Toeplitz operator for S^2}
	\mel*{i}{T_{p}(f)}{j}
	=\frac{p}{2\pi}\int_{S^2}\omega
	\braket*{i}{z}f(z,\bar z)\braket*{z}{i}.
\eeq
Let us focus on the embedding coordinates $x^A$
from  $S^2$ to $\textbf{R}^3$,
which are smooth real valued functions on $S^2$.
From (\ref{stereographic}), we have
\als
{
	\label{embedding coordinates}
	x^1&=\frac{z+\bar z}{1+|z|^2},\\
	x^2&=\frac{\im(\bar z-z)}{1+|z|^2},\\
	x^3&=\frac{1-|z|^2}{1+|z|^2}.
}
It is easy to find that the Toeplitz operators of $x^A$ are given by
\beq
	T_p(x^A)=\frac{L^A}{J+1},
\label{X for fuzzy s2}
\eeq
where $L^A$ are the $p$-dimensional irreducible representation of
the generators of $SU(2)$. 
This is the well-known configuration of the fuzzy $S^2$.

\subsection{Holomorphic matrix diffeomorphisms}

Here, we consider the matrix diffeomorphisms (\ref{diff of Mat}) 
for $X^A:=T_p(x^A)$. Since there are infinitely many 
diffeomorphisms even for the simple manifold $S^2$, we 
restrict ourselves to the holomorphic diffeomorphisms 
$\varphi\in\text{Aut}\,(S^2,J)$ in the following.
See appendix~\ref{appendix automorphisms} for a review of 
some automorphisms on $S^2$.

As reviewed in appendix~\ref{appendix automorphisms},
any $\varphi\in\text{Aut}\,(S^2,J)$ is expressed as
a M$\ddot{\text{o}}$bius transformation 
(\ref{Mobius}).
We focus on the four special transformations,
\als
{R_\theta(z)&=e^{\im\theta}z,\\ 
D_\lambda(z)&=e^{\lambda}z,\\ 
T_\eta(z)&=z+\eta,\\
S_\zeta(z)&=\frac{z}{\zeta z+1},
}
where $\theta,\lambda\in\mathbf{R}$ and $\eta,\zeta\in\mathbf{C}$.
These are a rotation, a dilatation, a translation and 
a special conformal transformation, respectively.
Note that any M$\ddot{\text{o}}$bius transformation can be
constructed as their composition\footnote{
In fact, for $c=0$, the M$\ddot{\text{o}}$bius transformation is
linear and is given by
a composition of $R_\theta$, $D_\lambda$ and $T_\eta$.
For $c\neq0$, it is expressed as
$\varphi(z)=(T_{(a-1)/c}\circ S_c\circ T_{(d-1)/c})(z)$.
}.
Note also that $R_\theta$ is an automorphism of $(S^2,\omega,J,g)$
satisfying the condition (\ref{condition for preserve g 2'}),
while the other three transformations are not.
We consider one-parameter groups, 
$\{R_{t\theta}\}_{t\in\textbf{R}}$, $\{D_{t\lambda}\}_{t\in\textbf{R}}$,
$\{T_{t\eta}\}_{t\in\textbf{R}}$ and $\{S_{t\zeta}\}_{t\in\textbf{R}}$,
which generate the vector fields defined by (\ref{diff vector}),
\als
{
	\label{four vector}
	u_R&=\im\theta(z\partial_z-\bar z\partial_{\bar z}),\\
	u_D&=\lambda (z\partial_z+\bar z\partial_{\bar z}),\\
	u_T&=\eta\partial_z+\bar \eta\partial_{\bar z},\\
	u_S&=-\zeta z^2\partial_z-\bar \zeta\bar z^2\partial_{\bar z},
}
respectively.

For a diffeomorphism generated by a vector field $u$, 
the infinitesimal variation of the embedding function $x^A$ is given 
as the Lie derivative $L_{u}x^A$, as reviewed in section
\ref{section diff}.
Correspondingly, the variation of the matrices are given 
by $\delta X^A=T_p(L_{u} x^A)$.
Let $X^\pm=T_p(x^\pm)= T_p(x^1\pm\im x^2)$. 
After some calculations, we easily find that
the infinitesimal variations of
$X^A$ for the vector fields (\ref{four vector})
 are given by
\begin{align}
	\begin{split}
	\label{delta_R}
	&\delta_RX^+=i\theta X^+,\\
	&\delta_RX^-=-i\theta X^-,\\
	&\delta_RX^3=0,\\
	\end{split}
	\\ \nonumber
	\\[-16pt]
	\begin{split}
	&\delta_DX^+=\lambda X^3X^++O(p^{-1}),\\
	&\delta_DX^-=\lambda X^3X^-+O(p^{-1}),\\
	&\delta_DX^3=-\lambda X^+X^-+O(p^{-1}),\\
	\end{split}
	\\ \nonumber
	\\[-16pt]
	\begin{split}
	&\delta_TX^+=\frac{1}{2}\eta(\textbf{1}_p+X^3)^2
	-\frac{1}{2}\bar \eta(X^+)^2+O(p^{-1}),\\
	&\delta_TX^-=\frac{1}{2}\bar \eta(\textbf{1}_p+X^3)^2
	-\frac{1}{2}\eta(X^-)^2+O(p^{-1}),\\
	&\delta_TX^3=-\frac{1}{2}(\textbf{1}_p+X^3)(\bar \eta X^++\eta X^-)+O(p^{-1}),\\
	\end{split}
	\\ \nonumber
	\\[-16pt]
	\begin{split}
	&\delta_SX^+=\frac{1}{2}\bar \zeta(\textbf{1}_p-X^3)^2
	-\frac{1}{2}\zeta(X^+)^2+O(p^{-1}),\\
	&\delta_SX^-=\frac{1}{2}\zeta(\textbf{1}_p-X^3)^2
	-\frac{1}{2}\bar \zeta(X^-)^2+O(p^{-1}),\\
	&\delta_SX^3=\frac{1}{2}(\textbf{1}_p-X^3)(\zeta X^++\bar \zeta X^-)
	+O(p^{-1}).
	\end{split}
\end{align}
The rotation (\ref{delta_R}) can be written as
$\delta_RX^A=-\im p[X^A,\theta X^3 / 2]+O(p^{-1})$.
This is the infinitesimal transformation of a unitary similarity 
transformation.
More generally, we show in 
appendix \ref{Aut(S^2,omega,J,g)} that 
any matrix diffeomorphism
corresponding to $\varphi\in\text{Aut}(S^2,\omega,J,g)$ is
given by a unitary similarity transformation.

We also notice that the other three matrix diffeomorphisms are
not unitary similarity transformations.
For example, let us check the case of $\delta_DX^A$.
If $\delta_DX^3$ is a similarity transformation,
we have 
$\delta_DX^3 \propto [U,X^3]$ with $U$ a certain matrix. 
Then, we will have
\beq
	\label{necessary condition for similarity}
	\ev*{\delta_DX^3}{Jr}=0,
\eeq
for all $r$. 
However,
$\ev*{\delta_D X^3}{Jr}=-\lambda(J+r)(J-r+1)/(J+1)^2$ is not zero for $r\neq-J$.
Thus, the matrix diffeomorphism corresponding to $D_{t\lambda}$ 
is not a similarity transformation.

Our definition of the matrix diffeomorphisms also works for 
finite transformations.
As an example, let us consider the dilatation. 
The finite diffeomorphism transforms of
(\ref{embedding coordinates}) are given by
\als
{
	D^\ast_{t\lambda }x^1&=\frac{e^{t\lambda}(z+\bar z)}{1+e^{2t\lambda}|z|^2},\\
	D^\ast_{t\lambda }x^2&=\frac{ie^{t\lambda}(\bar z-z)}{1+e^{2t\lambda}|z|^2},\\
	D^\ast_{t\lambda }x^3&=\frac{1-e^{2t\lambda}|z|^2}{1+e^{2t\lambda}|z|^2}.
}
In the following, we set $\lambda=1$ and $t\geq0$ for simplicity.
For example, the matrix elements $\mel*{J r}{T_p(D^\ast_{t}x^3)}{Jr'}$ reduces to
the following integral,
\als
{
	I
	:=\int_{S^2}\omega\,\frac{z^{J-r}\bar z^{J-r'}}{(1+|z|^2)^{2J}}
	\frac{1-e^{2t}|z|^2}{1+e^{2t}|z|^2}. \\
}
After integrating out the argument of $z$
and exchanging the integral variable from $|z|^2$ to $y=1/(1+|z|^2)$,
we obtain
\als
{
	\label{int for X'^3}
	I
	&=2\pi\delta_{rr'}(1+e^{-2t})
	\int^1_0dy\,y^{J+r+1}(1-y)^{J-r}\{1-(1-e^{-2t})y\}^{-1}\\
	&\qquad-2\pi\delta_{rr'}\int^1_0dy\,y^{J+r}(1-y)^{J-r}\{1-(1-e^{-2t})y\}^{-1}.
}
For a while, we suppose that $t\neq0$.
For $t>0$, we have $|1-e^{-2t}|<1$.
Using the integral representation of Gauss's hyper geometric function
$F(\alpha,\beta,\gamma;s)$ for $|s|<1$ and $0<\alpha<\gamma$,
\beq
	F(\alpha,\beta,\gamma;s)
	=\frac{\Gamma(\gamma)}{\Gamma(\alpha)\Gamma(\gamma-\alpha)}
	\int^1_0dy\,y^{\alpha-1}(1-y)^{\gamma-\alpha-1}(1-sy)^{-\beta},
\eeq
we can rewrite (\ref{int for X'^3}) as
\als
{
	I&=2\pi\delta_{rr'}(1+e^{-2t})\frac{\Gamma(J+r+2)\Gamma(J-r+1)}{\Gamma(2J+3)}
	F(J+r+2,1,2J+3;1-e^{-2t})\\
	&\qquad-2\pi\delta_{rr'}\frac{\Gamma(J+r+1)\Gamma(J-r+1)}{\Gamma(2J+2)}
	F(J+r+1,1,2J+2;1-e^{-2t}).
}
The calculations of the Toeplitz operators for 
$D^\ast_{t}x^+$ and $D^\ast_{t}x^-$ also reduce to similar integral
problems.
After evaluating the integrals, we find that the matrix elements 
of $T_p(D^\ast_{t}x^A)$ are given as
\als
{
	\label{dilatation of X}
	\mel*{J r}{T_p(D^\ast_{t}x^+)}{Jr'}
	&=\delta_{r-1 r'} \frac{e^{-t}}{J+1}\sqrt{(J- r+1)(J+ r)}
	F(J+r+1,1,2J+3;1-e^{-2t}),\\
	\mel*{J r}{T_p(D^\ast_{t}x^-)}{Jr'}
	&=\delta_{r+1 r'} \frac{e^{-t}}{J+1}\sqrt{(J+ r+1)(J- r)}
	F(J+r+2,1,2J+3;1-e^{-2t}),\\
	\mel*{J r}{T_p(D^\ast_{t}x^3)}{Jr'}
	&=\delta_{rr'}\frac{1}{2(J+1)}\{(1+e^{-2t})(J+r+1)F(J+r+2,1,2J+3;1-e^{-2t})\\
	&\qquad\qquad-2(J+1)F(J+r+1,1,2J+2;1-e^{-2t})\}.
}
Since $F(\alpha,\beta,\gamma:0)=1$, we have $T_p(D^\ast_{0}x^A)=X^A$.
Thus, the supposition of $t\neq0$ can be removed.

Again, we check that
$T_p(D^\ast_{t}x^A)$ is not related to $X^A$
by a unitary similarity transformation.
In the left figure of Fig. 2, we can see that 
the eigenvalue set of $T_p(D^\ast_{t}x^3)$ 
for $t=0.4$ is clearly different from the original 
eigenvalue set of $X^3$.
This shows that the map $X^A\mapsto T_p(D^\ast_{t}x^A)$ is not
a unitary similarity transformation.

The Toeplitz operators $X^A$ satisfy
\begin{align}
	\sum_{A=1}^3 X^AX^A=\textbf{1}_{p}+O(p^{-1}),
	\label{xx=1 for fuzzy sphere}
\end{align}
corresponding to the constraint $\sum_A x^Ax^A=1$.
Since any diffeomorphism does not break this constraint,
the matrix diffeomorphism $X^A \mapsto
T_p(D^\ast_{t}x^A)$ should also keep the equation
(\ref{xx=1 for fuzzy sphere}).
We check this as follows. The matrix
$\sum_A(T_p(D^\ast_{t}x^A))^2$ is diagonal and  
the eigenvalues are given by
\als
{
	\label{Square for dilatation}
	&\sum_{A=1}^{3}\ev*{(T_p(D^\ast_{t}x^A))^2}{Jr}\\
	&\qquad=\frac{1}{4(J+1)^2}\{(1+e^{-2t})(J+r+1)F(J+r+2,1,2J+3;1-e^{-2t})\\
	&\qquad\qquad\quad-2(J+1)F(J+r+1,1,2J+2;1-e^{-2t})\}^2\\
	&\qquad\qquad+\frac{e^{-2t}}{2(J+1)^2}
	\{(J-r+1)(J+r)F(J+r+1,1,2J+3;1-e^{-2t})^2\\
	&\qquad\qquad\quad+(J+r+1)(J-r)F(J+r+2,1,2J+3;1-e^{-2t})^2\}.
}
The right figure of Fig. 2 shows the plot of (\ref{Square for dilatation})
for $J=100000$ and $t=0.4$. 
Obviously, all the eigenvalues are equal to $1$.
Hence, the relation (\ref{xx=1 for fuzzy sphere}) also holds for the 
diffeomorphism transforms.

\begin{figure}
	\centering
	\includegraphics[width=16.5cm]{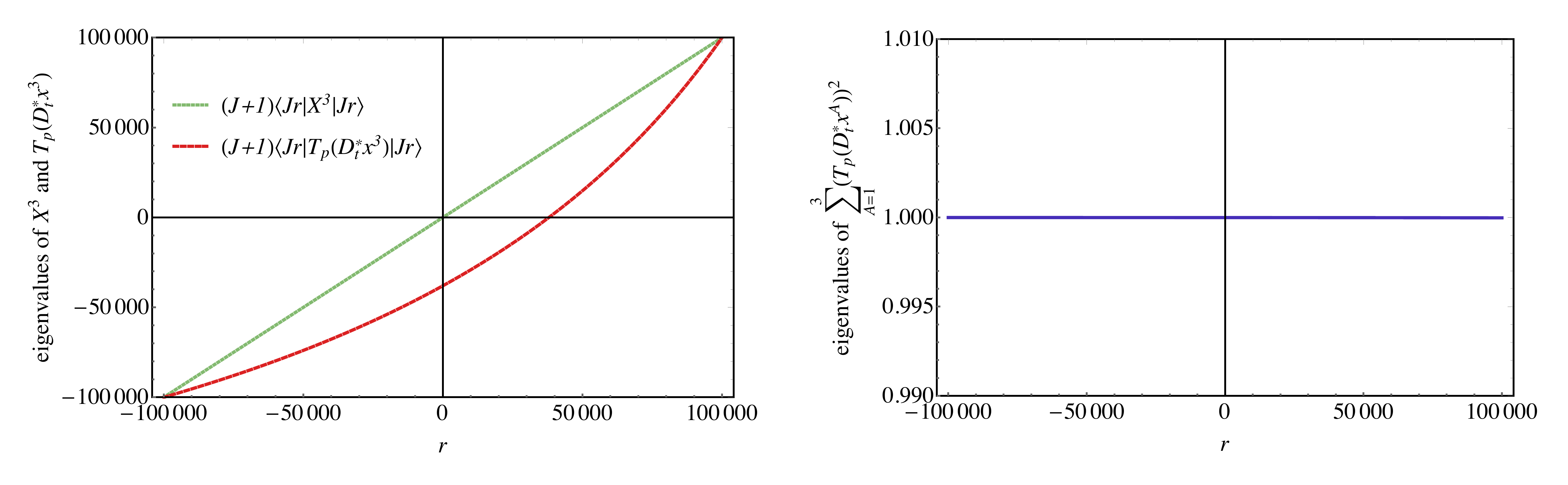}
	\caption
	{The green dotted line, the red dashed line and the blue solid line
	show the eigenvalues of $(J+1)X^3$, $(J+1)T_p(D^\ast_{t}x^3)$
	and $\sum_A(T_p(D^\ast_{t}x^A))^2$ for $J=100000$ and $t=0.4$,
	respectively.}
\end{figure}

\section{Approximate diffeomorphism invariants}

In this section, we propose three kinds of approximate 
invariants for the matrix diffeomorphisms on the fuzzy $S^2$.
These are functions $I(X)$ of the Toeplitz operators $X^A=T_p(x^A)$
which satisfy
\beq
	\label{relation for approximate invariants}
	I(X+\delta X)=I(X)+O(p^{-1}),
\eeq
for any infinitesimal matrix diffeomorphism $\delta X$
on the fuzzy $S^2$. In particular, if $\delta X$ is an infinitesimal
unitary transformation, then they satisfy $I(X+\delta X)=I(X)$.

\subsection{Invariants from matrix Dirac operator}
\label{section invariant 1}
For $p\times p$ matrices $X^A$ $(A=1,2,3)$ 
and the embedding function $x^A$ defined in (\ref{embedding coordinates}),
let us define a Dirac type operator,
\begin{align}
	\hat{D}= \sum_{A=1}^3\sigma^A \otimes (X^A -\hat x^A).
\label{matrix dirac op}
\end{align}
Here, we put a hat on $x^A$ to emphasize that $\hat x^A$ are 
kept fixed when we discuss the variation of approximate 
invariants, (\ref{relation for approximate invariants})
($\hat x^A$ are equal to $x^A$ as functions, $\hat x^A = x^A$.).
We also introduce the eigenstates of $\hat{D}$ as
\begin{align}
	\hat{D}\ket*{n} = E_n \ket*{n},
\end{align}
where the eigenvalues shall be labeled such that 
$|E_0| \leq |E_1| \leq |E_2| \leq \cdots$.
Note that $\hat{D}$, $\ket*{n}$ and $E_n$ depend on
local coordinates on $S^2$ through $\hat x^A$, although the 
dependences are not written explicitly. 
Apart from the fixed embedding function, the 
operator (\ref{matrix dirac op}) depends only on the matrices $X^A$.
In this sense, $E_n$ and $\ket*{n}$ are functions of $X_A$. 
The eigenvalues $E_n$ are not invariant for general transformations 
of matrices $X^A\mapsto X^{'A}$, but are exactly invariant under 
the unitary similarity transformations.

In the following, we consider the case in which 
$X^A$ are given by the Toeplitz operators of the embedding 
function (\ref{embedding coordinates}).
By solving the eigenvalue problem for this case
\cite{deBadyn:2015sca,Asakawa:2018gxf,Ishiki:2018aja}, 
one can find that $E_0$ and
$\ket*{0}$ are given by 
\als
{
	&E_0 = \frac{J}{J+1}-1 = O(p^{-1}), \\
	&|0 \rangle = U_2
	\left(
	\begin{array}{c}
	1 \\
	0 \\
	\end{array}
	\right)
	\otimes \ket*{z}.
}
Here, $U_2=e^{z\sigma^-}e^{-\sigma^3\text{log}(1+|z|^2)}e^{-\bar z\sigma^+}$
is a local rotation matrix and $\ket*{z}$ is the 
Bloch coherent state (\ref{def of bloch}).

The eigenvalue $E_0$, which has the smallest absolute value, 
gives our first example of the approximate invariants. 
Under an infinitesimal variation $X^A \mapsto X^A +\delta X^A$, 
$E_0$ transforms as\footnote{
This is just the first order formula of the perturbation theory 
in quantum mechanics. 
}
\begin{align}
	\delta E_0 = \sum^{3}_{A=1} \mel*{0}{\sigma^A \otimes \delta X^A}{0}.
\label{deltaE0}
\end{align}
We again emphasize that here $\hat{x}^A$ are kept fixed and 
we consider only the variation of the matrices.
Now, suppose that $\delta X^A$ is given by a 
matrix diffeomorphism, which can be written as
\begin{align}
	\delta X^A = \frac{p}{2\pi} \int_{S^2} \omega
	\ket*{w} \delta x^A(w) \bra*{w},
\end{align}
where $\delta x^A $ is the variation
of $x^A$ under a diffeomorphism.
Then, (\ref{deltaE0}) is evaluated as
\begin{align}
	\delta E_0 = \sum^{3}_{A=1} x^A \delta x^A +O(p^{-1}).
	\label{deltaE02}
\end{align} 
In deriving (\ref{deltaE02}), the following property of 
the Bloch coherent state is useful:
\als
{
	|\braket*{z}{w}|^2
	&= \frac{|1+w\bar{z}|^{4J}}{(1+|z|^2)^{2J}(1+|w|^2)^{2J}}, \\
	&= \exp 
	\left[
	2J \log 
	\left\{ 
	1-\frac{|z-w|^2}{(1+|z|^2)(1+|w|^2)}
	\right\}
	\right],\\
	& = \frac{\pi}{2J}(1+|z|^2)^2 \delta^{(2)}(z-w) +O(p^{-2}).
}
Since $\sum_{A}x^Ax^A=1$, the first term of (\ref{deltaE02}) is 
vanishing. Thus, $E_0$ is indeed 
invariant under the matrix diffeomorphism up to the $1/p$ corrections.

In \cite{Berenstein:2012ts}, 
it was proposed that the matrix Dirac operator can be used to 
find effective shapes of fuzzy branes.
Here, the loci of the zero eigenvalue of the matrix Dirac operator 
are identified with the effective shape embedded in the flat 
target space. See also \cite{deBadyn:2015sca,Karczmarek:2015gda}.
The same method was also independently proposed in the 
context of the tachyon condensation in string theory
\cite{Terashima:2005ic,Terashima:2018tyi,Asakawa:2018gxf}.

In \cite{Ishiki:2015saa,Ishiki:2016yjp,Schneiderbauer:2016wub}, 
to extract the classical shape of noncommutative spaces, 
another operator $\hat{H}=\sum_A(X^A-\hat x^A)^2/2$ was considered. 
For matrices which become commuting in the limit of large matrix size, 
$\hat{H}$ is equivalent to $\hat{D}^2$. Thus, the ground state energy 
of $\hat{H}$ also gives an approximate invariant of the matrix diffeomorphisms.

These invariants have 
the information of the induced metric for the embedding $\hat x^A$. 
As shown in \cite{Ishiki:2015saa}, by considering variations of $\hat{x}^A$,
we can construct from $E_0$ the Levi-Civita connection and
the Riemann curvature tensor for the induced metric.

\subsection{Invariants of information metric}
\label{section invariant 2}
In the space of density matrices, one can define the information metric,
\begin{align}
	ds^2 = \Tr (d\rho G),  \quad d\rho = \rho G + G \rho,
	\label{inf met}
\end{align}
where $\rho$ is a density matrix and $G$ is determined from $\rho$ 
by the second equation.
One can also restrict oneself to pure states  
$\{\rho = \op*{\psi}  
\mid \braket*{\psi} =1\}$.
In this case, $G=d\rho$ and the metric 
(\ref{inf met}) is equivalent to the Fubini-Study metric 
in the space of all normalized vectors
$\{\ket*{\psi} \mid \op*{\psi} =1 \}$, which has the 
structure of the complex projective space.

By using the eigenstate $\ket*{0}$ defined in the previous subsection, 
let us introduce a density matrix,
\begin{align}
	\rho = \op*{0}.
\end{align}
This gives an embedding of $S^2$ into the space of density matrices
\cite{Ishiki:2018aja}.
Then, the pullback $h$ of the information metric,
\begin{align}
	h_{\mu \nu}d\sigma^\mu d\sigma^\nu = \Tr d\rho d\rho,
\end{align}
gives a metric structure on $S^2$.

In our setup, the definition of $h$ depends on the choice of $X^A$ and $\hat x^A$.
However, in the setup of \cite{Berenstein:2012ts}, 
$\hat x^A$ are just thought of as three real 
parameters and the structure of embedding appears after solving the 
eigenvalue problem. The underlying space can be defined as the loci of 
zeros of the matrix Dirac operator. In this sense, the definition of 
$h$ depends only on the matrices $X^A$ and it gives a good geometric 
object defined in terms of the matrix variables.

Note that $h$ is exactly invariant under 
unitary similarity transformations $X^A \mapsto U^\dagger X^A U$.
Below, we show that the information metric is also approximately covariant 
under general matrix diffeomorphisms.
First, because $E_0 \rightarrow 0$ $(p \rightarrow \infty)$, 
we have $\ev*{\hat{D}^2}{0} \rightarrow 0$. 
This implies that $(X^A-\hat x^A)\ket*{0} \rightarrow 0$ for $A=1,2,3$. 
Let $\delta x^A$ be a polynomial of $x^A$ with the degree much 
less than $p$. Then, we also have 
\begin{align}
	(\delta X^A-\delta x^A)\ket*{0} \rightarrow 0  
	\label{X-x goes to zero}
\end{align}
as $p \rightarrow \infty$, where $\delta X^A$ 
is the Toeplitz operator of $\delta x^A$.
Let $\delta x^A$ be a Lie derivative of $x^A$ and 
$\delta X^A$ the corresponding matrix diffeomorphism.
Under the matrix diffeomorphism $X^A\mapsto X^A+ \delta X^A$, 
the state $|0\rangle $ transforms as 
\als
{
	\delta |0 \rangle
	&= \sum_{n\neq 0} \sum^{3}_{A=1}
	\frac{\op*{n} \sigma^A \otimes \delta X^A\ket*{0}}{E_0 -E_n}
	+\im\delta \lambda \ket*{0}, \\
	&= \sum_{n\neq 0} \sum^{3}_{A=1}
	\frac{\op*{n} \sigma^A  \ket*{0} \delta x^A }{E_0 -E_n}
	+\im\delta \lambda \ket*{0} + O(p^{-1}),
	\label{delta zero ket}
}
where $\delta \lambda$ is a real number and we used 
 (\ref{X-x goes to zero}) to obtain the last expression.
 We again emphasize that we fix $\hat x^A$ and consider only the variation of
 $X^A$.
On the other hand, from the infinitesimal variation of the local coordinates, 
we obtain
\als
{
	\partial_\mu |0 \rangle =  -
	\sum_{n\neq 0} \sum^{3}_{A=1}
	\frac{\op*{n} \sigma^A \ket{0}  \partial_{\mu}x^A }{E_0 -E_n}
	+\im A_\mu \ket*{0},
	\label{partial zero ket}
}
where $A = -\im\ev*{d}{0}$ is the Berry connection.
For a diffeomorphism $\delta x^A = u^\mu \partial_\mu x^A$,
from (\ref{delta zero ket}) and (\ref{partial zero ket}), 
we find 
\begin{align}
\delta \rho = -u^\mu \partial_\mu \rho+O(p^{-1}).
\end{align}
This means that the embedding function $\rho$ transforms
as a scalar field
under matrix diffeomorphisms.
Thus, the induced metric $h$ is also covariant:
\begin{align}
\delta h_{\mu \nu} = - \nabla_\mu u_\nu -\nabla_\nu u_\mu
+O(p^{-1}).
\end{align}

Diffeomorphism invariants (in the usual sense) 
defined in terms of $h$ are also approximately invariant 
under matrix diffeomorphisms. For example, 
the volume integral $\int_{S^2}\sqrt{h}$ or 
the Einstein-Hilbert action $\int_{S^2}\sqrt{h}R$ gives 
an approximate invariant.

In general, the information metric is different from the 
induced metric discussed in the previous subsection. 
For K$\ddot{\text{a}}$hler manifolds, 
the information metric gives a K$\ddot{\text{a}}$hler metric compatible 
with the field strength of the Berry connection
\cite{Ishiki:2016yjp}.
Hence, it has intrinsic information on the manifold, which 
does not depend on the embedding.

\subsection{Heat kernel on fuzzy sphere}
For a $2n$-dimensional closed Riemannian manifold $(M,g)$, the heat kernel, 
\begin{align}
	K(t)= \Tr e^{-t \Delta},
	\label{heat kernel}
\end{align}
for the Laplacian, $\Delta = -(1/\sqrt{g})\partial_{\mu}(\sqrt{g}
g^{\mu \nu}\partial_{\nu}) $, generates diffeomorphism invariants 
on $M$ as coefficients of the asymptotic expansion in $t\rightarrow +0$:
\begin{align}
	K(t)= \frac{1}{(4\pi t)^n }\int_M \sqrt{g}
	+ \frac{1}{(4\pi)^n t^{n-1} }\int_M \sqrt{g} \frac{R}{6} + \cdots.
\end{align}
Similarly, we define the heat kernel on the fuzzy $S^2$ as
\begin{align}
	\hat{K}(t_p, p)= \Tr e^{-t_p \hat{\Delta}}.
	\label{matrix heat kernel}
\end{align}
Here, $\hat{\Delta}$ is the matrix version of the Laplacian defined by
\begin{align}
	\label{matrix laplacian}
	\hat{\Delta} = (J+1)^2\sum_{A=1}^3 [X^A, [X^A,\,\cdot\,]]
	=\sum_{A=1}^3 [L^A, [L^A,\,\cdot\,]],
\end{align}
where $X^A=T_p(x^A)$ is given in (\ref{X for fuzzy s2}).
See \cite{Sasakura:2004dq,Sasakura:2005px} for 
the properties of $\hat{K}$ for finite size matrices.

It is well-known that the spectrum of $\hat{\Delta}$ coincides with 
that of the standard Laplacian on $S^2$ 
up to a UV cutoff given by the matrix size. 
The eigenstates of $\hat{\Delta}$ are given by the fuzzy spherical 
harmonics $\hat{Y}_{lm}$
\cite{Madore:1991bw,Grosse:1995jt,Baez:1998he,
Dasgupta:2002hx,Ishiki:2006yr,Ishii:2008ib}.
See appendix \ref{Fuzzy spherical harmonics} for the definition of 
$\hat{Y}_{lm}$, that we use in the following.
For $\hat{Y}_{lm}$, $l$ runs from $0$ to $p-1$ and 
$m$ runs from $-l$ to $l$. The eigenvalue of $\hat{\Delta}$ is 
$l(l+1)$ for $\hat{Y}_{lm}$, which coincides with 
the spectrum of the spherical harmonics on $S^2$, except that 
the angular momentum $l$ has a cutoff $p-1$ for the fuzzy spherical harmonics.

For finite $p$, the spectrum of $\hat\Delta$ is finite. 
Thus, the matrix heat kernel (\ref{matrix heat kernel}) has only a 
regular expansion in $t_p \rightarrow +0$ as 
$\hat{K}= \Tr \textbf{1}_{p^2} + O(t_p)$, which looks trivial 
and seems not to have any interesting information of the geometry.
However, it is obvious that if we first take the large-$p$ limit 
and then take $t_p \rightarrow +0$, $\hat{K}$ should 
behave similarly to $K$ having a singular expansion.
In other words, by putting $t_p=p^{-\alpha}$, where 
$\alpha$ is a small positive number, the heat kernel 
should have the expansion,
\begin{align}
\hat{K}(t_p=p^{-\alpha}, p) = \frac{1}{t_p}c_0 + c_1 + O(t_p)
\label{expansion of K hat}
\end{align}
in the large-$p$ limit.
It follows from the Euler-Maclaurin formula that
the coefficients are given by $c_0=1$ and $c_1=1/3$
for the Laplacian (\ref{matrix laplacian}).
See Fig. 3 for the plot of (\ref{expansion of K hat}).
The values of $c_0$ and $c_1$ just coincide with the coefficients 
of the heat kernel expansion on the continuum $S^2$.
Thus, in the double scaling limit, the matrix heat kernel 
possesses geometric information of $S^2$.

\begin{figure}
	\centering
	\includegraphics[width=8cm]{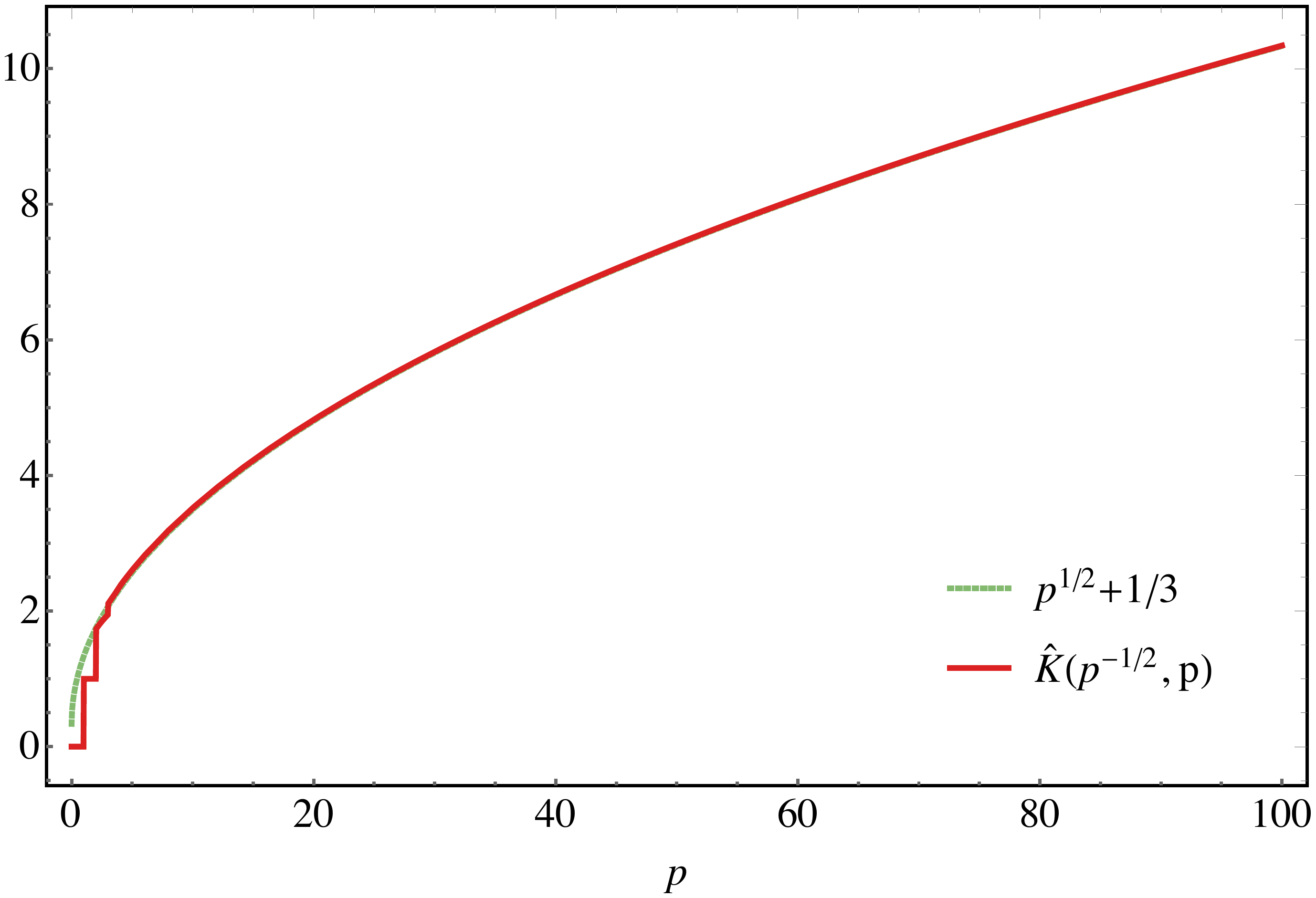}
	\caption
	{The green dotted line and the red solid line show $p^{1/2}+1/3$
	and $\hat K$ with $t_p=p^{-1/2}$, respectively.}
\end{figure}

Now, we show that the matrix heat kernel 
(\ref{matrix heat kernel}) is approximately 
invariant under matrix diffeomorphisms. 
Let us consider a perturbation 
$X^A \mapsto X^A + \delta X^A$.
Let $\delta X^A$ be a general 
infinitesimal matrix for the moment.
(In the end of the calculation, 
we will restrict $\delta X^A$ to be a matrix diffeomorphism.)
The eigenvalues of $\hat{\Delta}$ are perturbed by $\delta X^A$. 
Let $\delta_{lm}$ be the deviation of the 
eigenvalue for the mode $\hat{Y}_{lm}$. 
From the first order formula of the perturbation theory,
one obtains that
\begin{align}
	\delta_{lm}=\frac{(J+1)}{p}
	\Tr \sum^{3}_{A=1}\left(
	\hat{Y}^\dagger_{lm}
	[\delta X^A, [L^A, \hat{Y}_{lm}]]
	+\hat{Y}^\dagger_{lm}
	[L^A, [\delta X^A, \hat{Y}_{lm}]]
	\right).
	\label{delta lm}
\end{align}
The heat kernel (\ref{matrix heat kernel}) changes by 
\begin{align}
	\delta \hat{K} = -t_p
	\sum_{l=0}^{p-1}
	\sum_{m=-l}^l e^{-t_p l(l+1)}\delta_{lm}.
	\label{delta k hat}
\end{align}
The matrix $\delta X^A$ can be expanded in terms of 
the vector fuzzy spherical harmonics as 
\begin{align}
	\delta X^A
	=\sum^{p-1}_{l=0}\sum^l_{m=-l}\sum^1_{\rho=-1}
	\delta X_{lm\rho}\hat{Y}_{lm\rho}^A.
\end{align}
Again, see appendix \ref{Fuzzy spherical harmonics} 
for the definition of $\hat{Y}_{lm\rho}^A$.
After an easy calculation, we find that 
(\ref{delta k hat}) is given as 
\begin{align}
	\delta \hat{K} = 
	2\im t_p \delta X_{00-1}\sqrt{\frac{J+1}{J}}
	\sum_{l=0}^{p-1}e^{-t_p l(l+1)} l(l+1)(2l+1).
\end{align}
The important point is that the $\delta \hat{K}$ depends 
only on $\delta X_{00-1}$. This is exactly 
the mode proportional to $L^A$\footnote{Namely,  
if we consider a perturbation such that
$\delta X_{lm\rho}\propto \delta_{l0}\delta_{m0}\delta_{\rho -1}$,
such $\delta X^A$ is proportional to $L^A$.}.
This mode changes the radius of $S^2$ in the target space, 
and $\sum_A (X^A+\delta X^A)^2$ will deviate from the identity matrix even 
in the large-$p$ limit. 
Here, recall that, as mentioned in the previous section, 
any matrix diffeomorphism should keep the relation 
(\ref{xx=1 for fuzzy sphere}). 
The fluctuation of $\delta X_{00-1}$ violates this constraint, so it 
is not a matrix diffeomorphism. 
Therefore, for matrix diffeomorphisms, the matrix heat kernel is 
invariant. The coefficients in the expansion 
(\ref{expansion of K hat}) give approximate invariants on fuzzy $S^2$.

The matrix Laplacian corresponds to the operator 
$-\sum_A \{x^A, \{ x^A,\,\cdot\,\}\}$, because of (\ref{MR}).
This operator can be written as 
$-g^{\nu \sigma}
\partial_{\nu} \partial_{\sigma} + \cdots$, where 
$g^{\nu \sigma}= 
W^{\mu \nu}W^{\rho \sigma}\sum_A(\partial_{\mu}x^A\partial_{\rho}x^A)$
and $W^{\mu \nu}$ is the Poisson tensor.
The (inverse) metric $g^{\nu \sigma}$
is the open string metric \cite{Seiberg:1999vs} in the strong 
magnetic flux. 
Thus, the invariants from the heat kernel are associated with 
the open string metric.

\section{Summary and Discussion}

In this paper, we defined the action of diffeomorphisms on
the space of matrices through the matrix regularization.
We first constructed the matrix regularization of
closed symplectic manifolds based on the Berezin-Toeplitz quantization.
We then defined the matrix diffeomorphisms as
the matrix regularization of usual diffeomorphisms, as shown in Fig.~1.
We finally studied the matrix diffeomorphisms on the fuzzy $S^2$ and 
explicitly constructed holomorphic matrix diffeomorphisms.
We also constructed three kinds of approximate invariants of
the matrix diffeomorphisms on the fuzzy $S^2$. 
They are associated with three different kinds of metrics, the induced metric,
the K$\ddot{\text{a}}$hler metric and the open string metric.
In the case of $S^2$, they are equivalent up to an overall factor.
However, this is not the case for general spaces as shown in
\cite{Ishiki:2016yjp, Ishiki:2018aja}.
For example, it is easy to see this inequivalence by adding a perturbation 
to the fuzzy sphere.

The Berezin-Toeplitz quantization gives a systematic construction of
the matrix regularization for any compact symplectic manifold.
In the construction of Toeplitz operators that we discussed in this paper,
$\text{spin}^c$ structures play an essential role.
We emphasize that the existence of the symplectic structure is not essential in this construction. 
In fact, Toeplitz operators can also be constructed for $S^4$
\cite{Ishiki:2018aja,Zhang:2001xs,Hasebe:2010vp},
which is not a symplectic manifold.
Here, the well-known configuration of the fuzzy $S^4$ \cite{Castelino:1997rv} 
is obtained as the Toeplitz operator of the standard embedding function 
$S^4 \rightarrow \textbf{R}^5$.
It is known that any four dimensional oriented smooth manifold is 
a $\text{spin}^c$ manifold. Hence, Toeplitz operators can be constructed 
for any four-dimensional compact Riemannian manifolds.

In the matrix model formulation of M-theory, the fuzzy $S^4$ 
is interpreted as a longitudinal fivebrane \cite{Castelino:1997rv}. 
This example shows that the matrix model contains not only symplectic
manifolds but also more general manifolds with $\text{spin}^c$ structures. 
(Note that any D-brane must have a $\text{spin}^c$ structure.)
For general $\text{spin}^c$ manifolds without Poisson structure, 
the second condition in (\ref{MR}) for the matrix regularization 
can not be defined. 
However, the construction of Toeplitz operators is always 
possible and this may give a more fundamental framework 
of characterizing the matrix model.

Although we focused only on $S^2$ in this paper, 
our formulation can be straightforwardly extended to other spaces.
It will be important to study more general examples, in order to
understand the properties of matrix diffeomorphisms.
For example, the correspondence between area-preserving diffeomorphisms
and unitary similarity transformations may be more nontrivial for 
general cases.
When the first cohomology class is trivial, any area-preserving
diffeomorphism can be written in the form (\ref{area preserving for f}) 
and this is realized as a unitary similarity transformation 
in our definition of the matrix diffeomorphisms.
However, when the first cohomology class is nontrivial, there 
exist other area-preserving diffeomorphisms which cannot be written in 
the form (\ref{area preserving for f}).
It is interesting to study matrix diffeomorphisms corresponding to
such general area-preserving diffeomorphisms.

The approximate invariants we proposed in this paper 
are purely defined in terms of the matrix configuration of the fuzzy $S^2$. 
We consider that the constructions 
in section~\ref{section invariant 1} and \ref{section invariant 2}
can be generalized to the case of an arbitrary $\text{spin}^c$ manifold.
Such generalization may enable us to 
construct gravitational theories on fuzzy spaces. 
It is intriguing to pursue this direction.

\section*{Acknowledgments}
We thank N. Ishibashi and P. V. Nair 
for valuable discussions and encouraging comments.
The work of G. I. was supported, in part, 
by Program to Disseminate Tenure Tracking System, 
MEXT, Japan and by KAKENHI (16K17679).

\appendix 

\section{Berezin-Toeplitz quantization for classical mechanics}

In this appendix,  we consider the Berezin-Toeplitz quantization of 
a classical mechanical system of a particle on the real line.

We introduce a complex coordinate $z=(q+\im p)/\sqrt{2}$ for
the canonical variables $(q,p)\in\textbf{R}^2$.
We define a symplectic form on $\textbf{R}^2$ by
$\omega=dq\wedge dp=\im dz\wedge d\bar z$.
Then, the Poisson bracket defined by
$\omega$ satisfies $\{q,p\}=\im\{z,\bar z\}=1$.

Classical observables are just smooth functions 
on the phase space, $\{f(z,\bar{z}) \in C^{\infty}(\mathbf{R}^2) \}$.
The problem of the quantization is then to find a map from the 
classical observables to quantum observables $\{ \hat{f} \}$, which 
is a set of operators on a Hilbert space. It must be required that 
$\{f, g \}$ is mapped to $[ \hat{f}, \hat{g} ]/\im\hbar$ up to 
 higher order corrections of $\hbar$, where
$\hat{f}$ and $\hat{g}$ are the images of $f$ and $g$, respectively.
One can find such a map starting from the canonical operators 
$(\hat{q}, \hat{p})$ satisfying $[\hat{q}, \hat{p}]= \im\hbar $ and then 
fix the ordering of $(\hat{q}, \hat{p})$ in composite operators.

Each ordering gives a different quantization scheme. 
Among those, let us consider the anti-normal ordering. 
From $(\hat{q}, \hat{p})$, one can define the creation and 
annihilation operators 
$\hat{a}, \hat{a}^\dagger$ satisfying $[\hat a,\hat a^\dagger]=1$. 
In the anti-normal ordering,
$\hat{a}$ and $\hat{a}^\dagger$ are put on the left and right sides, respectively. 
Let $\ket{0}$ be the vacuum state defined by $\hat a\ket{0}=0$.
Then, the quantization map associated with this ordering can be written as 
\beq
	\label{BT map old}
	\hat{f}= T_{1/\hbar}(f)=\frac{1}{\pi\hbar}\int_{\mathbf{R}^2} \omega
	\ket*{z}f(z,\bar z)\bra*{z}
\eeq
for $f\in C^\infty(\textbf{R}^2)$,
where $\ket*{z}=e^{-|z|^2/2\hbar}e^{z \hat a^\dagger/\sqrt{\hbar}}\ket*{0}$ is the
canonical coherent state.
The overall factor $1/\pi\hbar$ is chosen such that $T_{1/\hbar}(1)=1$ holds. 
It is easy to check that this map satisfies the similar conditions to 
(\ref{MR})\footnote{The accuracy of the approximation in this case improves 
as $1/\hbar$ tends to infinity, i.e. in the classical limit.}.

There is a very useful reformulation of (\ref{BT map old}) in terms of 
Dirac zero modes. Let us consider the $U(1)$ gauge potential $A=(qdp-pdq)/2$ 
for the constant magnetic flux. The covariant Dirac operator is given by
\beq
	\label{Dirac op for R^2}
	D=\im\sigma^a\biggl(\partial_a-\frac{\im}{\hbar} A_a\biggr),
\eeq
where $\sigma^a$ $(a=1,2)$ is the Pauli matrix. 
The orthonormal basis of the Dirac zero modes is given by
\beq
	\label{LLL spinor}
	\psi_i(z,\bar z)= \frac{1}{\sqrt{\pi \hbar}}
	\left(\begin{array}{c}\braket*{i}{z}\\0\end{array}\right),
\eeq
where $\{ \ket{i}\mid i=1,2,\cdots \}$ is any orthonormal basis of the Hilbert space.
In terms of the zero modes (\ref{LLL spinor}), we can rewrite (\ref{BT map old}) as
\beq
	\label{BT map new}
	\mel{i}{T_{1/\hbar}(f)}{j}= \int_{\textbf{R}^2}\omega\,
	\psi^\dagger_j(z,\bar z)f(z,\bar z)\psi_i(z,\bar z)\,.
\eeq
Note that the coherent states in (\ref{BT map old}) are represented as 
the covariant spinors in (\ref{BT map new}).

The operator ${T_{1/\hbar}(f)}$ is 
called the Toeplitz operator of $f$. 
In the form of (\ref{BT map new}), the Toeplitz operator is given by 
the restriction of $f$ onto the space of the Dirac zero modes. 
The zero modes (\ref{LLL spinor}) are the wave functions in
the lowest Landau level of the Hamiltonian for
a charged particle moving in a constant magnetic field.
Thus, one can also say that the Toeplitz operator is the restriction of 
functions onto the space of the lowest Landau level.

The basic data required for constructing (\ref{BT map new}) are 
the Riemannian metric, the $U(1)$ gauge field and the Dirac zero modes.
A big advantage for using spinors is that the same construction works 
also for more general manifolds.

\section{Estimation of $\text{dim}\text{Ker}D$}
\label{Estimation of dimkerD}

In this appendix, we show that the sequence $N_p,N_{p+1},\ldots$ 
defined in section~\ref{section 3} is strictly 
monotonically increasing for large $p$.

Let us define the chirality operator
$\gamma_{2n+1}=(-\im)^n\gamma_1\gamma_2\cdots\gamma_{2n}$.
Since $\gamma_{2n+1}$ is Hermitian and satisfies 
$\gamma_{2n+1}^2=\textbf{1}_{2^n}$,
we can decompose $W$ into the direct sum of
the eigenspaces $W^\pm$ with the eigenvalues $\pm1$.
Correspondingly, we have the decomposition
$S\otimes L_p=(S^+\otimes L_p)\oplus(S^-\otimes L_p)$
where $S^\pm$ are the sub bundles of $S$ with fibers $W^{\pm}$.
Since $D\gamma_{2n+1}=-\gamma_{2n+1}D$,
we have $D\psi\in\Gamma(S^\mp\otimes L_p)$ for
$\psi\in\Gamma(S^\pm\otimes L_p)$. Hence, $D$ has the form
\beq
	\label{decomposition of D}
	D=\left(\begin{array}{cc}0&D^-\\D^+&0\end{array}\right),
\eeq
where $D^\pm$ are the restrictions of $D$ to $\Gamma(S^\pm\otimes L_p)$.
We define the subspaces of $\text{Ker}D$ by
$\text{Ker}D^\pm=\text{Ker}D\cap\Gamma(S^\pm\otimes L_p)$.
Since (\ref{decomposition of D}) implies that
$\text{dim}\text{Ker}D=\text{dim}\text{Ker}D^++\text{dim}\text{Ker}D^-$,
we have
\beq
	\text{dim}\text{Ker}D\geq|\text{ind}D|,
\label{dimkerD equal indD}
\eeq
where $\text{ind}D=\text{dim}\text{Ker}D^+-\text{dim}\text{Ker}D^-$ is the
index of $D$. The equal sign holds if and only if 
$\text{dim}\text{Ker}D^+=0$ or $\text{dim}\text{Ker}D^-=0$. 
In addition, $\text{Ker}D^-=\{0\}$ holds 
in our setting for large $p$ 
because of the vanishing theorem 
\cite{Ma:2002}, so that we have 
\beq
\text{dim}\text{Ker}D=|\text{ind}D|. 
\label{dimker equal index}
\eeq
Moreover, the Atiyah-Singer index theorem gives a relation,
\beq
	\label{index theorem}
	\text{ind}D=\int_M \hat A(M)\wedge\text{ch}(L_p),
\eeq
where $\hat A(M)$ denotes the $\hat A$-genus of $M$
and $\text{ch}(L_p)$ the Chern character of $L_p$.
Then, we have the formula,
\beq
	\text{ch}(L_p)
	=(\text{ch}(L_1))^p
	=\text{exp}\biggl(\frac{ pF}{2\pi}\biggr),
\eeq
where the product of differential forms is defined by the wedge product.
From the assumption that $F/2\pi=\omega V_n^{-1/n}$, we find
\beq
	\text{ind}D=p^n +O(p^{n-2}).
\label{large p expansion of indD}
\eeq
From (\ref{dimker equal index}) and (\ref{large p expansion of indD}),
we conclude that $\{ N_p=\text{dim}\text{Ker}D \mid p\gg 1 \}$  
is indeed a strictly monotonically increasing sequence.

\section{Geometric structures on $S^2$}
\label{appendix geometric structure}

In this appendix, we review our notation for the geometry of 
$S^2$ and introduce some geometric structures.

Let $x^A$ $(A=1,2,3)$ be the Cartesian coordinates on $\mathbf{R}^3$.
We consider a two-dimensional unit sphere $S^2$
defined by the equation $\sum_{A=1}^3x^Ax^A=1$.
We identify $S^2$ with the Riemann sphere
$\hat{\mathbf{C}}=\mathbf{C}\cup\{\infty\}$
by the stereographic projection 
$S^2\to\hat{\mathbf{C}}$ defined by
\beq
	\label{stereographic}
	z =\frac{x^1+\im x^2}{1+x^3}
\eeq
for $x^3\neq-1$ and $z=\infty$ for $x^3=-1$.
Under this identification,
we can cover $S^2$ by two open subsets
$U_z:=\hat{\mathbf{C}}-\{\infty\}$
and $U_w:=\hat{\mathbf{C}}-\{0\}$.
Then, the coordinate neighborhood system of $S^2$
consists of $(U_z;z)$ and $(U_w;w:=1/z)$.
The coordinate transformation from $(U_z;z)$ to $(U_w;w)$
is given by a holomorphic map $z\mapsto1/z$.

The sphere $S^2$ is a K$\ddot{\text{a}}$hler manifold and
we can define a symplectic structure $\omega$, complex structure $J$
and Riemann structure $g$ such that 
they satisfy the compatible condition.
First, we define $\omega$ by a volume form on $S^2$,
\beq
 	\label{symplectic for S^2}
	\omega=\im\frac{dz\wedge d\bar z}{(1+|z|^2)^2},
\eeq
such that $\int_{S^2}\omega=2\pi$.
Secondly, we define $J$ by $J(\partial_z)=\im\partial_z$
and $J(\partial_{\bar z})=-\im\partial_{\bar z}$.
The local form is
\beq
	\label{almost complex for S^2}
	J=\im\partial_z\otimes dz-\im\partial_{\bar z}\otimes d\bar z.
\eeq
Finally, we define $g$ by the compatible condition 
$g(\,\cdot\,,\,\cdot\,)=\omega(\,\cdot\,,J\,\cdot\,)$ as
\beq
	\label{metric for S^2}
	g=2\frac{dzd\bar z}{(1+|z|^2)^2}.
\eeq

We choose an orthonormal frame on $U_z$
with respect to the metric (\ref{metric for S^2}) as
\als
{
	\label{orthonormal frame}
	e_1&=\frac{1}{\sqrt{2}}(1+|z|^2)(\partial_z+\partial_{\bar z}), \\
	e_2&=\frac{\im}{\sqrt{2}}(1+|z|^2)(\partial_z-\partial_{\bar z}).
}
Then, the dual basis is
\als
{
	\label{dual frame}
	\theta^1&=\frac{1}{\sqrt{2}}\frac{dz+d\bar z}{1+|z|^2}, \\
	\theta^2&=\frac{1}{\sqrt{2}\im}\frac{dz-d\bar z}{1+|z|^2}.
}
The linear map (\ref{linear map of e}) is given by
\beq
	\gamma(e_a)=\sigma_a,
\eeq
where $\sigma_a$ are Pauli matrices.
In this choice, the chirality operator is
$\Gamma=-\im \sigma_1\sigma_2=\sigma_3$.
The condition (\ref{spin connection}),
which determines the spin connection on $S$,
is equivalent to
\als
{
	&\Omega^1{}_2\wedge \theta^2-\frac{\im}{\sqrt{2}}(z-\bar z)
	\theta^1\wedge \theta^2=0, \\
	&\Omega^1{}_2\wedge \theta^1-\frac{1}{\sqrt{2}}(z+\bar z)
	\theta^2\wedge \theta^1=0.
}
By solving these equation, we obtain
\beq
	\Omega^1{}_2
	=\frac{\im}{\sqrt{2}}(z-\bar z)\theta^1+\frac{1}{\sqrt{2}}(z+\bar z)\theta^2
	=-\im\frac{\bar zdz-zd\bar z}{1+|z|^2}.
\eeq

We also need a topologically nontrivial configuration 
of the $U(1)$ gauge connection on $S^2$ to construct 
Toeplitz operators. We use the Wu-Yang monopole configuration,
\beq
	A^{(z)}=-\frac{\im}{2}\frac{\bar zdz-zd\bar z}{1+|z|^2},
\label{Wu-Yang}
\eeq
for $U_z$. On the overlap of two patches,
the gauge connection $A^{(w)}$ on $U_w$ is related 
to (\ref{Wu-Yang}) by a $U(1)$ gauge transformation.
More specifically, 
$A^{(w)}=A^{(z)}-d\,\text{arg}(z)$ on $U_z\cap U_w$. 
This gauge connection satisfies $F=dA^{(z)}=2\pi\omega V_1^{-1}$.

\section{Automorphisms on $S^2$}
\label{appendix automorphisms}

In this appendix, we review $\text{Aut}(S^2,J)$, 
$\text{Aut}(S^2,g)$ and $\text{Aut}(S^2,\omega)$.
See appendix~\ref{appendix geometric structure} for the definitions 
of $J$, $g$ and $\omega$.

\subsection{$\text{Aut}(S^2,J)$}
First, we consider $\text{Aut}(S^2,J)$.
For $\varphi\in\text{Diff}(S^2)$,
let $\hat z$ be a point on $S^2$ such that $\varphi(\hat z)=\infty$.
Namely, $\hat z$ is a pole of $\varphi$. 
Note that since $\varphi$ needs to be one-to-one, $\hat z$ is 
the unique pole. For simplicity, we first suppose that $\hat z=\infty$.
In this case, we have $\varphi(U_z)\subset U_z$.
The local form of the new tensor field $J'$
induced by $\varphi$ is given on $U_z$ as
\beq
	J'
	=\im\varphi^{-1}_\ast(\partial_z)\otimes \varphi^\ast(dz)
	-\im\varphi^{-1}_\ast(\partial_{\bar z})\otimes \varphi^\ast(d\bar z),
\eeq
where $\varphi_\ast$ is the pushforward by $\varphi$.
If $J'=J$, then we have
\als
{
	\label{condition for preserve J}
	&\partial_\varphi z\,\partial_z \varphi
	-\partial_{\bar\varphi} z\,\partial_z \bar\varphi=1,\\
	&\partial_\varphi\bar z\,\partial_z \varphi
	-\partial_{\bar\varphi}\bar z\,\partial_z \bar\varphi=0,
}
where $\partial_\varphi=\partial/\partial\varphi(z)$.
Note that $\varphi(z)$ generally depends on both $z$ and $\bar z$.
From the chain rule, $1=\partial_z z= \partial_\varphi z\,\partial_z \varphi
+\partial_{\bar\varphi}z\,\partial_z\bar\varphi$, and 
the first equation of (\ref{condition for preserve J}),
the relation $\partial_{\bar\varphi}z\,\partial_z\bar\varphi=0$ follows.
This shows that $\varphi(z)$ and $\varphi^{-1}(z)$ are holomorphic on $U_z$.
The second equation of (\ref{condition for preserve J}) automatically holds
when $\partial_z \bar\varphi=\partial_{\bar\varphi} z=0$.
In the case that $\hat z\neq\infty$,
a similar argument leads to the conclusion 
that $\varphi$ has a pole at $\hat z$
and is holomorphic at every points except at $\hat z$.
In summary, $\varphi$ preserving $J$ is
a meromorphic function on $S^2$
with a pole at a point.

One can express such $\varphi$
as $\varphi=f/h$, where $f$ and $h$ are relatively prime functions on $S^2$.
If the degree of $f$ or $h$ is second or higher,
$\varphi$ cannot be one-to-one.
Thus, both of $f$ and $h$ have to be at most linear polynomials 
and $\varphi \in \text{Aut}(S^2,J)$ is expressed as
\beq
	\label{Mobius}
	\varphi(z)=\frac{a z+b}{c z+d},
\eeq
where $a,b,c,d$ are complex numbers such that $ad-bc\neq0$\footnote{
The condition $ad-bc\neq0$ ensures that
$\varphi$ is not a constant function.
For $ad-bc=0$, we have $\varphi(z)=b/d$.}.
We define $\varphi(\infty)=\infty$ for $c=0$
and $\varphi(\infty)=a/c$ for $c\neq0$.
Since multiplying $a,b,c,d$ by a common number does not change
the value of (\ref{Mobius}), we can fix $ad-bc=1$.
This transformation is the so-called M$\ddot{\text{o}}$bius transformation
and the group $\text{Aut}(S^2,J)$ consists of
all M$\ddot{\text{o}}$bius transformations.

Let us consider 
a homomorphism $\Pi:SL(2,\mathbf{C})\to\text{Aut}(S^2,J)$ defined by
\beq
	\label{homomo}
	\left(\begin{array}{cc}a&b \\ c&d\end{array}\right)\mapsto\varphi.
\eeq
Then, we have $\text{Ker}\Pi=\{\pm \textbf{1}_2\}$.
From the fundamental theorem on homomorphisms,
we find that $\text{Aut}(S^2,J)$ is isomorphic to
$PSL(2,\mathbf{C})=SL(2,\mathbf{C})/\mathbf{Z}_2$.

\subsection{$\text{Aut}(S^2,g)$}
Secondly, we consider $\text{Aut}(S^2,g)$.
We suppose that $\hat z=\infty$ again.
The local form of the new tensor field $g'$
induced by $\varphi$ on $U_z$ is given by
\beq
	g'=2\frac{\varphi^\ast(dz)\varphi^\ast(d\bar z)}{(1+|\varphi(z)|^2)^2}.
\eeq
If $g'=g$, then we have
\als
{
	\label{condition for preserve metric}
	&\partial_{\bar z}\varphi\,\partial_{\bar z}\bar\varphi=0, \\
	&\partial_z\varphi\,\partial_{\bar z}\bar\varphi
	+\partial_{\bar z}\varphi\,\partial_z\bar\varphi
	=\frac{(1+|\varphi(z)|^2)^2}{(1+|z|^2)^2}.
}
From the first equation of (\ref{condition for preserve metric}),
$\partial_{\bar z}\varphi=0$ or $\partial_{\bar z}\bar\varphi=0$ follows.
The former and the later means that
$\varphi$ is holomorphic and anti-holomorphic on $U_z$, respectively.

In the case that $\varphi$ is holomorphic,
the same argument as $\text{Aut}(S^2,J)$ shows 
that $\varphi$ is given by
the M$\ddot{\text{o}}$bius transformation (\ref{Mobius}).
In the case that $\varphi$ is anti-holomorphic,
we can set $\varphi=\tilde\varphi\circ\theta$,
where $\theta\in\text{Diff}(S^2)$ is defined by $\theta(z)=\bar z$
and $\tilde\varphi\in\text{Diff}(S^2)$ is holomorphic on $U_z$.
Then, $\tilde\varphi$ is given by the M$\ddot{\text{o}}$bius 
transformation (\ref{Mobius}), so that $\varphi$ can be written as
\beq
	\label{anti Mobius tra}
	\varphi(z)=\frac{a \bar z+b}{c \bar z+d},
\eeq
where the definition of $\{a,b,c,d\}$ is the same as (\ref{Mobius}).
This transformation is called an anti-M$\ddot{\text{o}}$bius transformation.
The composition of two anti-M$\ddot{\text{o}}$bius transformations is
a M$\ddot{\text{o}}$bius transformation,
and the composition of a M$\ddot{\text{o}}$bius transformation
and an anti-M$\ddot{\text{o}}$bius transformation is
an anti-M$\ddot{\text{o}}$bius transformation.
Thus, all M$\ddot{\text{o}}$bius transformations
and anti-M$\ddot{\text{o}}$bius transformations form a group,
which is called the extended M$\ddot{\text{o}}$bius group
and denoted by $\overline{PSL}(2,\mathbf{C})$.

In any case, 
the second equation of (\ref{condition for preserve metric}) is equivalent to
\als
{
	\label{condition for preserve g 2'}
	|a|^2+|c|^2=|b|^2+|d|^2=1,\quad a\bar b+c\bar d=0.
}
This means that both $\Pi^{-1}(\varphi)$ and $\Pi^{-1}(\tilde\varphi)$
are elements of $PSU(2,\mathbf{C})=SU(2,\mathbf{C})/\vb*{Z}_2$.
We therefore find that $\text{Aut}(S^2,g)$ is isomorphic to
$\overline{PSU}(2,\mathbf{C})$
which is a subgroup of $\overline{PSL}(2,\mathbf{C})$
defined by the condition (\ref{condition for preserve g 2'}).
We also find that $\text{Aut}(S^2,J,g)$ is isomorphic to
$PSU(2,\mathbf{C})\cong SO(3)$.

Note that
$\text{Aut}(S^2,J,g)=\text{Aut}(S^2,\omega,J,g)$,
since $J$ and $g$ are compatible with $\omega$.

\subsection{$\text{Aut}(S^2,\omega)$}
Finally, we consider $\text{Aut}(S^2,\omega)$.
The local form of the new tensor field $\omega'$
induced by $\varphi$ on $U_z$ is given by
\beq
	\omega'=\im\frac{\varphi^\ast(dz)\wedge\varphi^\ast(d\bar z)}{(1+|\varphi(z)|^2)^2}.
\eeq
If $\omega'=\omega$, then we have
\beq
	\label{condition for preserve omega}
	\partial_z\varphi\,\partial_{\bar z}\bar\varphi
	-\partial_{\bar z}\varphi\,\partial_z\bar\varphi
	=\frac{(1+|\varphi(z)|^2)^2}{(1+|z|^2)^2}.
\eeq
Note that there is not an equation corresponding to the first equation of
(\ref{condition for preserve metric}).
We therefore cannot conclude that
$\varphi$ is holomorphic or anti-holomorphic on $U_z$.
This suggests that $\text{Aut}(S^2,\omega)$ is a larger group
than $\text{Aut}(S^2,J)$ and $\text{Aut}(S^2,g)$.
In fact, as reviewed in section~\ref{section diff}, 
the Lie algebra of $\text{Aut}(S^2,\omega)$ is isomorphic to
the Poisson algebra on $S^2$,
since the first cohomology class on $S^2$ is trivial.
If $\varphi$ is holomorphic,
satisfying (\ref{condition for preserve omega}) is equivalent to
$\varphi\in\text{Aut}(S^2,\omega,g,J)$.
If $\varphi$ is anti-holomorphic,
(\ref{condition for preserve omega}) never holds.
This corresponds to the fact that 
the orientation determined by $\omega$
is not kept under the inversions $z\mapsto \bar z$.

\section{Matrix diffeomorphisms for $\text{Aut}(S^2,\omega,J,g)$}
\label{Aut(S^2,omega,J,g)}

In this appendix, we show that matrix diffeomorphisms for
$\text{Aut}(S^2,\omega,J,g)$ can be written as 
unitary similarity transformations.

For any $\varphi \in \text{Aut}(S^2,\omega,J,g)$,
there exists an element $u\in SU(2,\mathbf{C})$ such that 
$\varphi=\Pi(u)$, where $\Pi$ is defined by (\ref{homomo}).
By using the relation of the stereographic coordinate (\ref{stereographic}),
it is easy to check that the following relation holds:
\beq
	\varphi^\ast x^A
	=\sum^3_{B=1}\Lambda^{AB}x^B,
\eeq
where $\Lambda\in SO(3)$ is the three dimensional irreducible 
representation of $u$.
There exists a unitary matrix $U$ (given by the $p$-dimensional representation 
of $u$) such that $\sum_B\Lambda^{AB}L^B=UL^AU^{-1}$. Hence, we find that
\beq
	\label{rotation vs unitary}
	\mel{i}{T_p(\varphi^\ast x^A)}{j}
	=\sum^3_{B=1}\Lambda^{AB}\mel*{i}{X^B}{j}
	=\mel*{i}{UX^AU^{-1}}{j}.
\eeq
In conclusion, any matrix diffeomorphism corresponding to
$\varphi\in\text{Aut}(S^2,J,g)$
is a unitary similarity transformation.

\section{Fuzzy spherical harmonics}
\label{Fuzzy spherical harmonics}

In this appendix, we review the definition of the fuzzy spherical harmonics
and the vector fuzzy spherical harmonics.
See \cite{Ishiki:2006yr,Ishii:2008ib} for more details.

The linear maps $[L^A,\,\cdot\,]$ on $M_{p}(\mathbf C)$
define a $p^2$-dimensional representation of the generators of
$SU(2)$ because they satisfy
$[[L^A,\,\cdot\,], [L^B,\,\cdot\,]]
=\im\sum^3_{C=1}\epsilon^{ABC}[L^C,\,\cdot\,]$.
The fuzzy spherical harmonics $\hat Y_{l m}$
$(l=0,1,\ldots,p-1, m=-l,-l+1,\ldots,l)$ are defined as the standard basis
of this representation space which satisfy
\als
{
	&[L^\pm, \hat Y_{l m}]
	=\sqrt{(l\mp m)(l\pm m+1)}\hat Y_{l m\pm1}, \\
	&[L^3, \hat Y_{l m}]
	=m\hat Y_{l m},
}
and the orthonormality condition
$\Tr \hat{Y}^\dagger_{lm}\hat{Y}_{l'm'}/p=\delta_{ll'}\delta_{mm'}$.
They are expressed in terms of the basis
$\{\op*{Jr}{Jr'}\mid r,r'=-J,-J+1,\ldots,J\}$ as
\beq
	\hat Y_{l m}
	=\sqrt{p}\sum^J_{r,r'=-J}(-1)^{-J+r'}C^{l m}_{JrJ-r'}\op*{Jr}{Jr'},
\eeq
where $C^{l m}_{JrJ-r'}$ is the Clebsch-Gordan coefficient.

The vector fuzzy spherical harmonics $\hat Y^A_{l m\rho}$
$(\rho=-1,0,1)$ are defined in terms of
the fuzzy spherical harmonics as
\beq
	\hat Y^A_{l m\rho}
	=\im^\rho\sum_{B=1}^{3}\sum_{n=-\tilde Q}^{\tilde Q}
	V^{AB}C^{Qm}_{\tilde Qn1B}\hat Y_{\tilde Qn},
\eeq
where $Q=l+\delta_{\rho1}$, $\tilde Q=l+\delta_{\rho -1}$
and $V$ is a unitary matrix given by
\beq
	V=\left(\begin{array}{ccc}-1&0&1\\-\im&0&-\im\\0&\sqrt{2}&0\end{array}\right).
\eeq
They also satisfy the orthonormality condition
$\sum^3_{A=1}\Tr \hat{Y}^{A\dagger}_{lm\rho}\hat{Y}^A_{l'm'\rho'}/p
=\delta_{ll'}\delta_{mm'}\delta_{\rho\rho'}$ and
transform as the vector representation under $SU(2)$ rotation.


\end{document}